\documentclass{aa}  

\usepackage{graphicx}
\usepackage{xspace}
\usepackage{txfonts}
\usepackage{subfig}
\usepackage{xcolor}
\usepackage{amsmath}
\usepackage{amsfonts} 

\begin{document}

   \title{The COSMOS Wall at $z \sim 0.73$: Quiescent galaxies and their evolution in different environments
\thanks{Based on observations made with ESO Telescopes at the European Southern Observatory, Cerro Paranal, Chile, under program IDs 194-A.2005 and 1100.A-0949 (The LEGA-C Public Spectroscopy Survey), and  ESO 085.A-0664 (The cosmos structure at z = 0.73: exploring the onset of environment-driven trends).}
}

 \author{F. R. Ditrani\inst{1,2}, M. Longhetti\inst{2}, A. Iovino\inst{2}, M. Fossati\inst{1,2}, S. Zhou\inst{2}, S. Bardelli\inst{3}, M. Bolzonella\inst{3}, O. Cucciati\inst{3}, A. Finoguenov\inst{4}, L. Pozzetti\inst{3}, M. Salvato\inst{5}, M. Scodeggio\inst{6}, L. Tasca\inst{7}, D. Vergani\inst{3}, and E. Zucca\inst{3}}
    \institute{Università degli studi di Milano-Bicocca, Piazza della scienza, I-20125 Milano, Italy\\
   \email{f.ditrani1@campus.unimib.it}
   \and
   INAF-Osservatorio Astronomico di Brera, via Brera 28, I-20121 Milano, Italy
   \and
   INAF - Osservatorio di Astrofisica e Scienza dello Spazio di Bologna, via Gobetti 93/3, I-40129 Bologna, Italy
   \and
   Department of Physics, University of Helsinki, Gustaf Hällströmin katu 2, 00560 Helsinki, Finland
   \and
   Max Planck Institute for Extraterrestrial Physics, Giessenbachstr. 1, 85748 Garching, Germany
   \and
   INAF - IASF Milano, Via Bassini 15, I-20133, Milano, Italy
   \and
   Laboratoire d’Astrophysique de Marseille, CNRS-Université d’Aix-Marseille, 38 rue F. Joliot Curie, F-13388 Marseille, France}

   \date{Received; accepted}
 \titlerunning {}
 \authorrunning{Ditrani et al.}
 
  \abstract
   {The evolution of quiescent galaxies is driven by numerous physical processes, often considered to be related to their stellar mass and environment over cosmic time. Tracing their stellar populations can provide insight into the processes that transformed these galaxies into their observed quiescent state. In particular, higher-redshift galaxies, being younger, exhibit more pronounced relative age differences. At early stages, even small differences in age remain significant, whereas as galaxies evolve, these differences become less detectable, making it harder to trace the impact of environmental effects in the local Universe.
   The COSMOS Wall is a structure at $z \sim 0.73$ that contains a large variety of environments, from rich and dense clusters down to field-like regions. Thus, this sample offers a great opportunity to study the effect of the environment on the quiescent galaxy population.
   

    
    Leveraging high-quality spectroscopic data from the LEGA-C survey, combined with the extensive photometric coverage of the COSMOS2020 catalogue, we performed a full-index and photometry fitting of $74$ massive ($\log{M_\star/M_\odot} = 10.47$) quiescent galaxies, deriving their mass-weighted ages, metallicities, and star formation timescales.
   We characterised the environment in three subsamples: X-ray- and non-X-ray-detected groups and a lower-density subsample similar to the average field.  
   We find a decreasing trend in mass-weighted age with increasing environmental density, with galaxies groups $\gtrsim 1$ Gyr older than those in the field. Conversely, we do not find any significant difference in stellar metallicity between galaxies in X-ray and non-X-ray groups, while we find galaxies with $0.15$ dex higher metallicities in the field. 

   Our results indicate that, at $z \sim 0.7$, the environment plays a crucial role in shaping the evolution of massive quiescent galaxies, noticeably affecting both their mass-weighted age and star formation timescale. These results support faster quenching mechanisms, at fixed stellar mass, in the dense X-ray-detected groups compared to the field.}
   
   
   


   \maketitle
%

\section{Introduction}

Determining the efficiency of and interplay between the different physical mechanisms leading to galaxy quenching is still one of the unsolved problems in galaxy evolution. Several processes can contribute to the suppression of star formation, and they can be grouped in two broad classes: those related to the galaxy properties (internal processes) and those related to the environment in which galaxies reside (external processes). The efficiency of internal processes increases with  galaxy mass, a key property that regulates the physics of mechanisms such as stellar and active galactic nucleus feedback \citep[e.g.][]{croton2006,fabian2012,cicone2014}. In contrast, the external processes involve a large variety of mechanisms, including ram-pressure stripping \citep[e.g.][]{gunn1972}, strangulation \citep[e.g.][]{larson1980,balogh2000}, and galaxy-galaxy interactions \citep[e.g.][]{mihos1996,moore1998}. Dynamical gas removal processes like ram-pressure stripping, typical in high-density environments \citep{boselli2022}, can rapidly deplete the cold gas reservoir, leading to a fast quenching event. Conversely, processes that reduce the cosmological gas inflow \citep[e.g. the `cosmological starvation' process presented by][]{man2018}, in combination with maintenance mode feedback processes \citep{dave2012}, can also occur in lower-density environments, or in the field, and potentially have longer quenching times \citep{man2018}. This scenario also results in an overall younger and more metal-enriched stellar population with respect to galaxies quenched by faster processes.
Although both mass and environmental quenching contribute to the suppression of star formation, their relative significance varies depending on the galaxy mass and the cosmic epoch. Moreover, the effects of these various quenching processes on galaxy properties are often degenerate, making it challenging to distinguish their individual contributions.

Quiescent galaxies are ideal laboratories for performing studies on which mechanisms lead to galaxy quenching since they contain fingerprints of the quenching processes. In particular, the physical properties of quiescent galaxies, such as the age and the metallicity of their stellar populations, can provide strong constraints on timescales and thus the physical origin of the quenching mechanisms.
Stellar population properties of quiescent galaxies have been known to vary strongly with the environment in which galaxies are located \citep[][]{spitzer1951,oemler1974,davis1976,dressler1980}. In dense environments and massive haloes, the evolution from star-forming galaxies to quiescent ones takes place sooner than in the field \citep{cooper2006,cucciati2006,poggianti2006,iovino2010,peng2010}.
However, whether this environmental signal is merely the result of different galaxy masses in different environments (i.e. the mass segregation framework) or reflects intrinsic differences in stellar history at fixed mass is still debated \cite[e.g.][]{baldry2006,thomas2010,poggianti2013}.
Studies on the effect of the environment are mainly performed in the context of the local Universe. In particular, the stellar ages and metallicities of quiescent galaxies have been shown to correlate to first order with galaxy mass \citep[e.g.][]{gallazzi2005ages,thomas2005epochs,thomas2010}, with environment playing a larger role in less massive galaxies \citep[e.g.][]{thomas2010,gallazzi2021galaxy}. Some studies have been able to determine the role of the environment in  quiescent galaxies and found older galaxies in the most massive environments with respect to the field \citep[][]{sanchez2006medium}.

Thanks to the high-quality spectroscopic data of The Large Early Galaxy Astrophysics Census (LEGA-C) Public Spectroscopic Survey \citep[][]{van2016vlt,straatman2018large,van2021large}, we are now able to study a large sample of quiescent galaxies at higher redshifts ($0.6 < z < 1$), where differences in stellar populations could be more pronounced than in the local Universe. The LEGA-C survey targeted galaxies in the Cosmological Evolution Survey (COSMOS) field \citep[][]{scoville2007}, in which a prominent large-scale structure was detected in a narrow redshift slice, $0.69 \le z \le 0.79$ \citep[][]{scoville2007b,cassata2007,guzzo2007}. This structure, named the COSMOS Wall \citep{Iovino2016}, provides a unique opportunity to investigate the effects of the environment on galaxy evolution. This volume covers a comprehensive range of environments, ranging from a dense cluster to filaments and voids. Moreover, the COSMOS Wall redshift, corresponding to roughly half the age of the Universe ($\sim 6.5$ Gyr), marks a transitional period when the Universe's star formation rate (SFR) was declining and environmental effects began to play a dominant role in shaping galaxy properties \citep{fossati2017}. For quiescent galaxies specifically, this epoch provides key insights into how and when they ceased star formation, as well as the mechanisms responsible for their transformation.

In this work we present our analysis of the stellar populations of quiescent galaxies in this peculiar field. 
We assembled a multi-wavelength dataset in order to exploit all the available information. We used a hierarchical Bayesian method to determine the average stellar population parameters for galaxies in different environments. A complementary study \citep{Zhou25} will focus on star-forming galaxies in the COSMOS-Wall volume for a parallel analysis. 

In Sect.~\ref{sec:wall} we present the full set of data used to perform our analysis and the definition of the different environments in the COSMOS Wall. In Sect.~\ref{sec:analysis} we describe in detail the procedures adopted to retrieve the stellar population parameters of our selected sample. In Sect.~\ref{sec:resultdisc} we present the results that we obtained and the discussion of our analysis.
Throughout the paper we adopt a standard $\Lambda$ cold dark matter cosmology with $\Omega_M = 0.286$, $\Omega_\Lambda = 0.714,$ and $H_0 = 69.6$ km s$^{-1}$ Mpc$^{-1}$ \citep{bennett20141}. Magnitudes are in the AB system \citep{oke1974absolute}.

\section{Data and sample selection} 
\label{sec:wall}

The COSMOS Wall is a volume identified within the COSMOS survey that contains a large variety of environments, from rich and dense clusters (with X-ray detection) to poor and loose groups, down to average field regions \citep{scoville2007b}. This volume is at redshift $0.69 \le z \le 0.79$, located in the RA-Dec region displayed in Fig. $1$ of \cite{Iovino2016}. Readers are referred to Iovino et al. (2016) for more details on this structure’s definition.
\cite{Iovino2016} performed a detailed mapping of the COSMOS Wall volume using a friends-of-friends algorithm and an iterative procedure to obtain a reliable group catalogue at different spatial scales. The valuable aspect of this structure consists in the opportunity it offers to study the possible differences in the physical properties of galaxies in relation to their environment, at a fixed cosmic epoch.

The COSMOS field has been targeted by several spectroscopic surveys \citep[e.g.][]{lilly2007zcosmos,coil2011,comparat2015}, and among them the LEGA-C survey \citep{van2016vlt} provided high signal to noise ratio and high resolution spectra of the more massive galaxies in this area. Indeed the spectra, observed with the Visible Multi-Object Spectrograph (VIMOS) on the Very Large Telescope (VLT), have an average continuum $S/N \sim 20 \ \AA^{-1}$, with an instrumental spectral resolution ($R \sim 3500$) suitable to study the stellar population properties of galaxies \citep[DR3;][]{van2021large}.
In this work we studied the physical properties of a sample of massive quiescent galaxies within the COSMOS Wall volume by combining optical LEGA-C spectra and a set of photometric data covering a large wavelength range. 
For the photometric data, we used the multi-wavelength photometric COSMOS$2020$ catalogue \citep{weaver2022}. In particular, we used the smallest photometric aperture magnitudes available ($2$ arcsec) measured in $u^*$ band from the MegaCam/CFHT images, in the HSC-SSP PDR2 $g$-, $r$-, $i$- ,$z$- and $y$-band images and in the UltraVISTA DR4 $J$-, $H$- and $K_s$-band \cite{weaver2022}, in order to match the $1$ arcsec slit aperture of the LEGA-C observations \citep{van2021large}.

From the entire LEGA-C sample, we selected galaxies that are in the COSMOS Wall volume as defined above, identifying a total of $244$ objects. We measured the available star formation indicators ([OII]$_{\lambda3727}$, H$\delta$, H$\beta$, [OIII]$_{\lambda5007}$) using the latest version of the  penalised pixel-fitting (pPXF) code \citep[][]{cappellari2004parametric,cappellari2017improving,cappellari2023}, adopting the E-MILES simple stellar population (SSP) synthesis models \citep{vazdekis2016uv}, obtained assuming the BaSTI tracks \citep{pietrinferni2004large,pietrinferni2006large} and \cite{chabrier2003galactic} initial mass function (IMF). We then spectroscopically selected a subsample of quiescent galaxy candidates using the following criteria:
\begin{itemize}
    \item EW([OIII]$_{\lambda5007}$) $< 1 \ \AA$;
    \item EW(H$\beta$) $< 1 \ \AA$;
    \item EW([OII]$_{\lambda3727}$) $< 5 \ \AA$;
    \item EW(H$\delta$) $< 3 \ \AA$ in absorption.
\end{itemize}
These criteria are designed to select galaxies with no detectable [OIII]$_{\lambda5007}$ and H$\beta$ emission lines and at most a weak [OII]$_{\lambda3727}$ emission, which can occur in quiescent galaxies at these redshifts 
\citep[see][]{maseda2021}. Additionally, the last criterion, EW(H$\delta$) $< 3 \ \AA$ in absorption, excludes post-starburst galaxies from our sample, following the definition of \cite{poggianti2009}. Using the above selection criteria, we defined a sample of $99$ quiescent galaxies.
However, due to the limited and variable rest-frame spectral range, approximately half of the selected sample lacks at least one of the  indicators required for this selection. Specifically, among the $99$ selected galaxies, $19$ have spectra not including the [OII] line, $14$ galaxies lack both [OIII] and H$\beta$, and $9$ ones lack only the [OIII]. For these galaxies, we based the selection on the other available indicators.  We tested the possible contamination introduced by using a reduced set of criteria using the subsample of $109$ out of $244$ galaxies for which all the four indicators are available. For example, we performed the selection without considering the [OII] line and then checked the values of [OII] for the selected galaxies. We found that some galaxies would not meet the selection criteria if the [OII] threshold of $5 \AA$ had been applied.
This analysis revealed that, in the worst-case scenario, only $5$ out of the $99$ galaxies in the sample might not satisfy all the selection criteria when the full set of indicators is considered.

\begin{figure}
   \centering
   \includegraphics[width=0.5\textwidth]{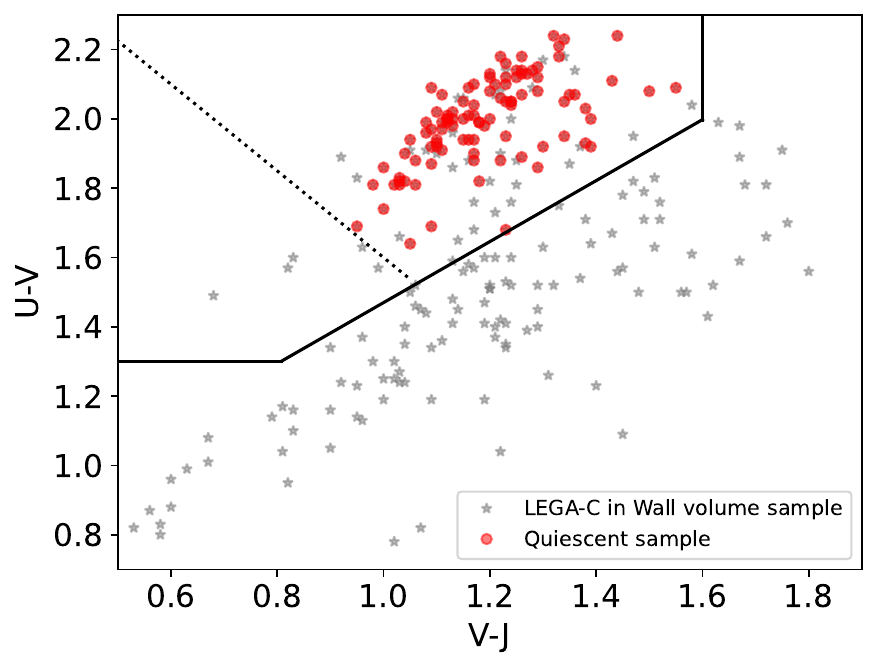}
      \caption{Rest-frame UVJ diagram of the selected sample. The stars are all the galaxies observed in the COSMOS Wall volume by the LEGA-C survey, while the red points mark the spectroscopically selected quiescent galaxies. The solid black line is the separation between star-forming and quiescent galaxies as defined in Eq. $4$ of \cite{williams2009}. The dotted line divides younger quiescent galaxies (on the left) and older quiescent galaxies (on the right) according to \cite{whitaker2013}.}
         \label{fig:nuvj_sampleandmass}
   \end{figure}

 Figure~\ref{fig:nuvj_sampleandmass} shows the rest-frame UVJ diagram of the $244$ galaxies observed in the COSMOS Wall volume by the LEGA-C survey (grey points) and our selected quiescent sample (red points). Rest-frame $U-V$ and $V-J$ values are obtained from the LEGA-C photometric parent sample \cite{muzzin2013}. The entire sample of selected galaxies falls within the region of the UVJ diagram associated with quiescent galaxies, as defined by \cite{williams2009}. Specifically, they occupy the region corresponding to  older quiescent galaxies, as defined from \cite{whitaker2013}, further supporting the robustness of our selection. We also inspected the spectra of those galaxies that have colours matching those of the quiescent galaxies in the UVJ plane but that were excluded by our spectroscopic selection. We found that their spectra clearly exhibit gas-phase emission lines in those galaxies. Therefore, we consider the spectroscopic criteria we adopted for selecting quiescent galaxies to be solid and trustworthy.
 
\cite{Iovino2016} classified galaxies in the COSMOS Wall volume as belonging to groups or as field galaxies. Furthermore, by cross-correlating their group catalogue with the list of XMM-COSMOS extended sources presented in \cite{george2011} (down to a limit X-ray detection from of rest-frame $\log(L_{X,0.1-2.4keV}/erg s^{-1}) \sim 41.3$, corresponding to $\log(M_{200}/M_\odot) > 13.2$), they further categorised the groups into X-ray emitters or non-X-ray emitters. This matching process identified nine groups in the Wall Volume with X-ray detection. Notably, the new X-ray group catalogue for the COSMOS field, derived by \cite{Gozaliasl2019}, using combined Chandra and XMM-Newton observations from the COSMOS Legacy Survey, does not alter the list of matched groups originally presented by \cite{Iovino2016}.
Our sample galaxies can then be divided into three different environment bins: groups with X-ray detection, groups without X-ray detection, and field.
Specifically, our sample consists of $22$ galaxies located in groups with X-ray detection (hereafter X-ray groups), $28$ in groups without X-ray detection (hereafter non-X-ray groups), and $49$ in the field.

\begin{figure}
   \centering
   \includegraphics[width=0.5\textwidth]{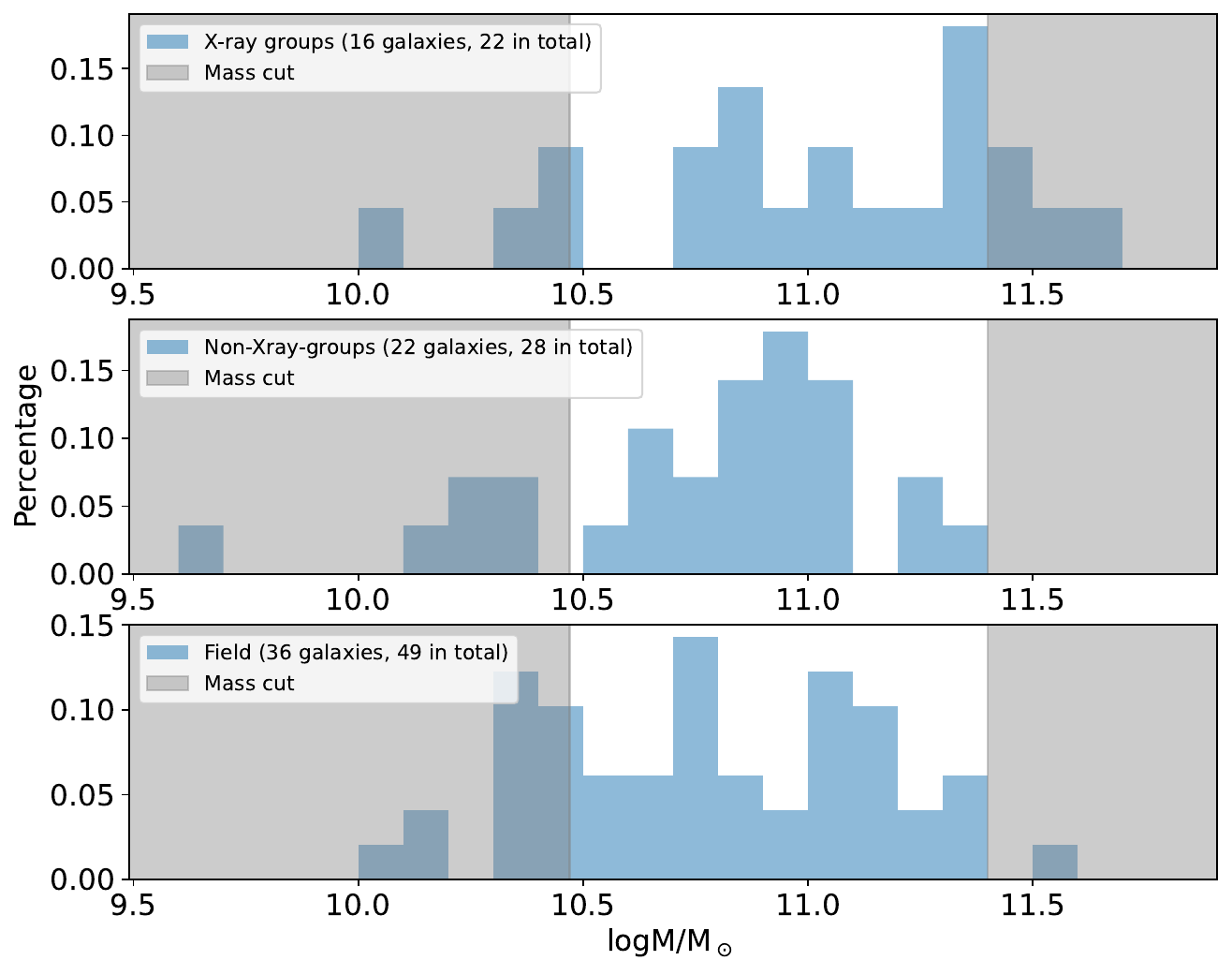}
      \caption{Mass distribution in the three environment bins. The shaded grey region is the mass cut we applied to obtain a homogeneous and representative distribution across the three bins.}
         \label{fig:mass_distrib_concut}
   \end{figure}

Figure~\ref{fig:mass_distrib_concut} shows the stellar mass distribution of the galaxies across the three environment bins. We used stellar masses estimated from \cite{muzzin2013}. It is well known that galaxy evolution strongly depends on their stellar mass \citep[e.g.][]{thomas2010}, and to isolate the effect of environment, we need to compare samples with the same mass distribution across the three environments. According to \cite{van2021large}, the LEGA-C sample in the redshift bin between $0.7$ and $0.8$ is representative down to a stellar mass limit of $\log{M/M_\odot} = 10.47$. Therefore, we refined our sample selection by restricting it to galaxies within the mass range $10.47 < \log{M/M_\odot} < 11.4$, since we used the same measurements from \cite{muzzin2013}. The upper limit was set to exclude the more massive galaxy tail, which is present only in the X-ray group environment.  We performed a K-S test for the three distribution in the restricted mass bin, and we obtained a p-value of $0.2$ between X-ray and non-X-ray groups, a p-value of $0.4$ between X-ray groups and field, and a p-value of $0.5$ between non-X-ray groups and field. From the K-S test we did not detect significant differences between the three population and therefore we assumed that the distributions are extracted from the same parent population. Our final sample consists then of $16$ galaxies located in X-ray groups, $22$ in non-X-ray groups, and $36$ in the field, for a total of $74$ quiescent galaxies. 


\section{Analysis}
Our goal is to measure the individual mass-weighted age, the stellar metallicity and the star formation timescale of our sample of quiescent galaxies, and then investigate the effects of the environment on their properties.
In this section we describe in detail the steps of the analysis on our sample. Specifically, 
in Sect.~\ref{subsec:models} we introduce the synthetic templates built to analyse the selected galaxies. In Sect.~\ref{subsec:jointphotandspec} we describe the performed joint fitting of spectra and photometry using Monte Carlo techniques, given the multi-wavelength available data (Sect.~\ref{subsec:jointphotandspec}). Finally, we present how we assess the stellar population parameters of each environment bin using the hierarchical Bayesian modelling (Sect.~\ref{subsec:hierarc}).

\label{sec:analysis}
\subsection{Stellar population models}
\label{subsec:models}
We built a set of synthetic templates based on the SSP models from the latest version of the \citet[]{bruzual2003stellar} library \citep[hereafter the C\&B library; see][]{plat2019constraints,sanchez2022sdss}. The SSP models adopt the PARSEC evolutionary tracks \citep{marigo2013evolution,chen2015parsec}. We assumed a Chabrier IMF \citep[][]{chabrier2003galactic} with $M_{UP} = 100 M_\odot$ and the MILES stellar library \citep[$3540.5 < \lambda < 7350.2 \ \AA$;][$2.5\AA$ full width at half maximum resolution]{sanchez2006medium,falcon2011updated}. The C\&B library provides $3300$ SSPs unevenly spaced in linear age and [Z/H], covering $220$ ages from $0.01$ Myr to $14$ Gyr and $15$ metallicities from [Z/H]$\ = -2.23$ to [Z/H]$\ = 0.55$ dex; for reference, the solar abundance (Z$_\odot$) is $0.017$. 
From the C\&B library we generated composite stellar population templates characterised by a fast rise of star formation, typical of massive quiescent galaxies \citep[e.g.][]{citro2016}, followed by an exponentially declining star formation history (SFH):
\begin{equation}
   SFR_\tau(t) \sim \sqrt{\left(\frac{t}{\tau}\right)}\exp{\left(-\frac{t}{\tau}\right)} 
,\end{equation}
where $\tau$ represents the SFR timescale of the SFH, varying between $0.1 \le \tau \le 2$ Gyr, in increments of $0.1$ Gyr. At $t = 2\tau$ the SFR has declined to approximately $30\%$ of its peak value and formed about $75\%$ of the total stellar mass. Finally, we adopted the attenuation law from \cite{Calzetti:2001}  with $0 \le A_V \le 1$ mag. Given the age of the Universe at the redshift of the COSMOS Wall ($\sim 7$ Gyr), we set the maximum age of the templates to $8$ Gyr.

\subsection{Joint spectroscopic and photometric analysis}
\label{subsec:jointphotandspec}
In our analysis, we combined spectroscopic and photometric data to derive the stellar population parameters of the selected galaxies through comparisons with the spectral template library described in the previous subsection. 
For each galaxy in the sample, we determined the kinematic parameters (i.e. recession velocity and velocity dispersion) using the latest version of the pPXF code, adopting the template set from the MILES stellar spectral library \citep{sanchez2006medium}, convolved to match the LEGA-C instrumental resolution, and considering the rest-frame wavelength range between $3540 \ \AA$ and $4700 \ \AA$. Synthetic templates were then shifted to match the recession velocity of each galaxy and convolved with the estimated stellar velocity dispersion. Our fit of the velocity dispersion are consistent within $10$ km/s with the available values from \cite{van2021large}.

\begin{figure*}
   \centering
   \includegraphics[width=0.95\textwidth]{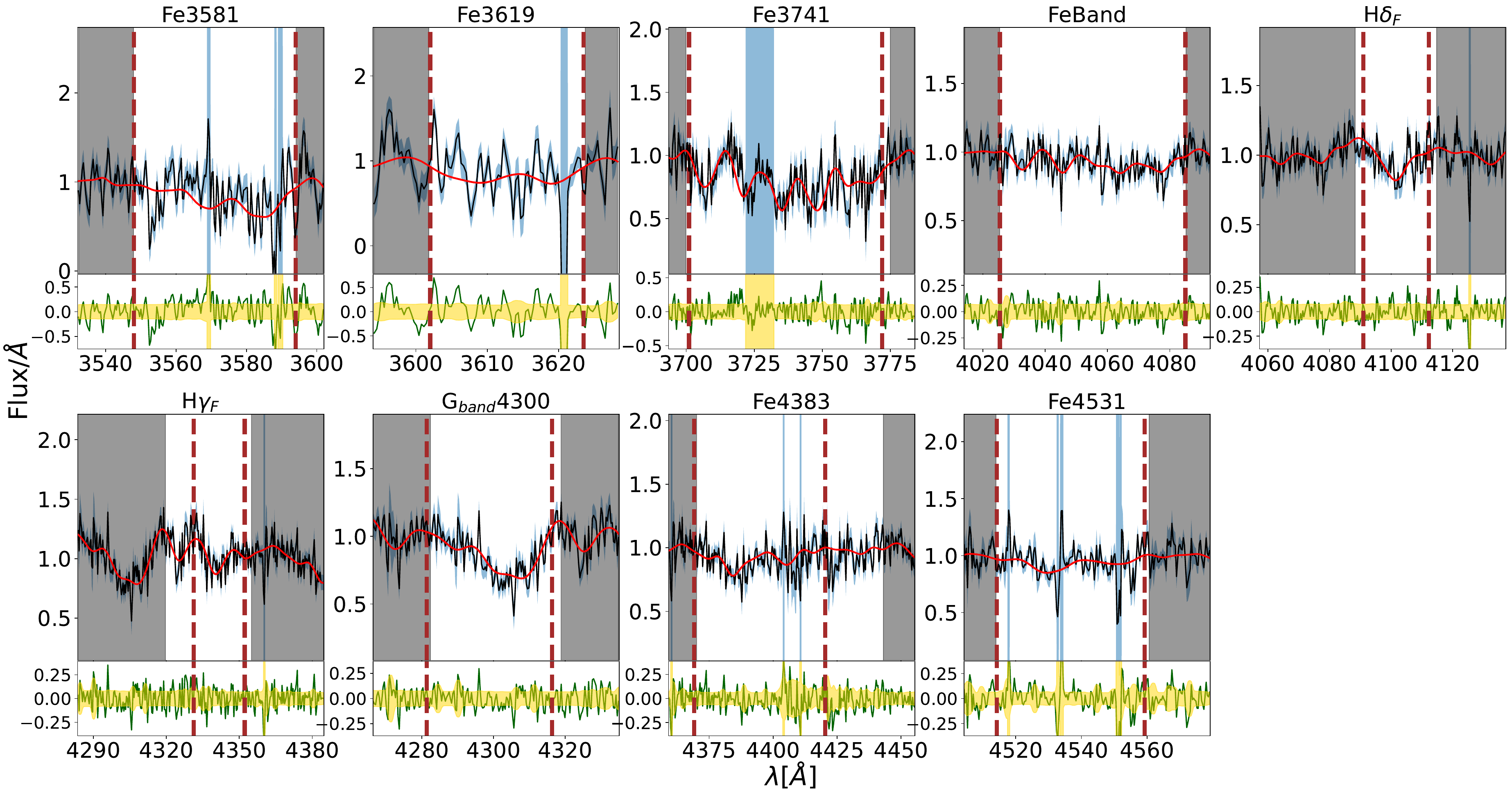}
      \caption{FIF application to the indices available in the LEGA-C ID 209466 galaxy spectrum. The vertical dashed brown lines indicate the feature boundaries for each index, while the grey shaded regions represent the pseudo-continua used for normalisation. In the upper subplots, the black lines and blue shaded regions correspond to the observed spectrum and its associated uncertainty, respectively. The solid red line represents the best-fit derived from the posterior distribution. The green lines in the lower subplots show the residuals between the observed spectrum and the best fit, with the yellow shaded region indicating the relative uncertainties of the observed spectrum.
              }
         \label{fig:confr_fif}
         
   \end{figure*}

We then adopted the full-index fitting (FIF) approach \citep[][]{martin2014stellar} to compare the observed spectra with the synthetic spectral templates. Differently from the more classic index fitting approach, FIF compares the flux within a specific absorption feature (with respect to its continuum value) pixel by pixel rather than averaging it.  This pixel-level comparison within the index window is more effective at breaking the degeneracy between age and metallicity compared to the classical index analysis, as it accounts not only for the strength of the absorption feature but also for its detailed shape, which provides additional information about the stellar population parameters. \citep[][]{martin2019fornax,ditrani24}. 
We applied the FIF method to a comprehensive set of spectral indices, namely: Fe$3581$, Fe$3619$, Fe$3741$, D$_{\text{n}}4000$, FeBand$4050$, H$\delta_F$, H$\gamma_F$, G$_{band}4300$, Fe$4383$, and Fe$4531$ \citep[][]{gregg1994spectrophotometry,balogh1999differential,Worthey1994old}. Given the available wavelength range, we used only Fe indicators to determine the stellar metallicity, meaning that the derived total metallicity is a proxy of the  [Fe/H] abundance.
Figure~\ref{fig:confr_fif} shows an example of the FIF approach applied to one of the selected galaxies. The best-fit template closely matches each spectral feature, capturing detailed information from both their depth and pixel-by-pixel shape.

The only exception to the FIF approach concerns the D$_\text{n}4000$ index. For this index, we adopted the classic index definition, and its value was compared with the corresponding values measured in the synthetic templates.

Regarding the photometric data, we compared the observed fluxes in the $9$ bands described in Sect.~\ref{sec:wall} with those measured in the same photometric bands on the synthetic spectral templates. For this comparison, we first normalised the synthetic templates using the observed $i$-band flux corresponding to the wavelength range of the LEGA-C observed spectrum. Additionally, we fitted three `jitter' parameters: $\ln f_{uv}$ for the $u^*$ band, $\ln f_{opt}$ for $g$-, $r$-, $i$- ,$z$- and $y$-band and $\ln f_{ir}$ for $J$-, $H$- and $K_s$-band. These parameters act as multiplicative factors applied to the photometric uncertainties in order to balance the likelihoods between the spectroscopic and the photometric fits, given that the S/N of the photometric data significantly exceeds that of the spectroscopic data. 
 
\begin{figure*}
   \centering
   \includegraphics[width=0.95\textwidth]{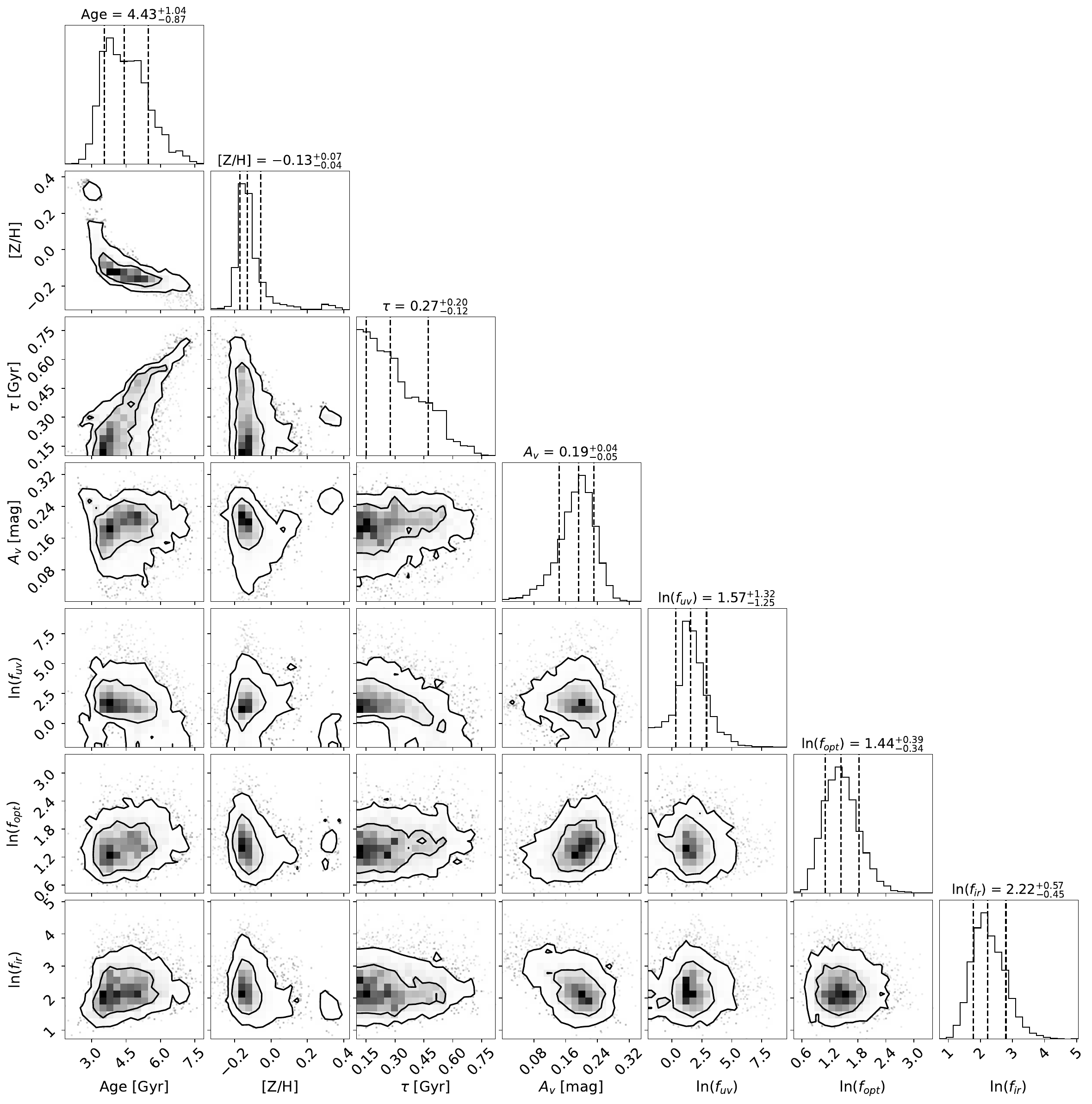}
      \caption{Example of the joint and marginal posterior distributions for the LEGA-C ID 209466 galaxy. Contours represent the $68\%$ and $95\%$ probability levels. The $16\%$, $50\%,$ and $84\%$ intervals are indicated as dashed lines.}
         \label{fig:corner_all}
   \end{figure*}

\begin{figure*}
   \centering
   \subfloat[][\emph{}]
        {\includegraphics[width=0.45\textwidth]{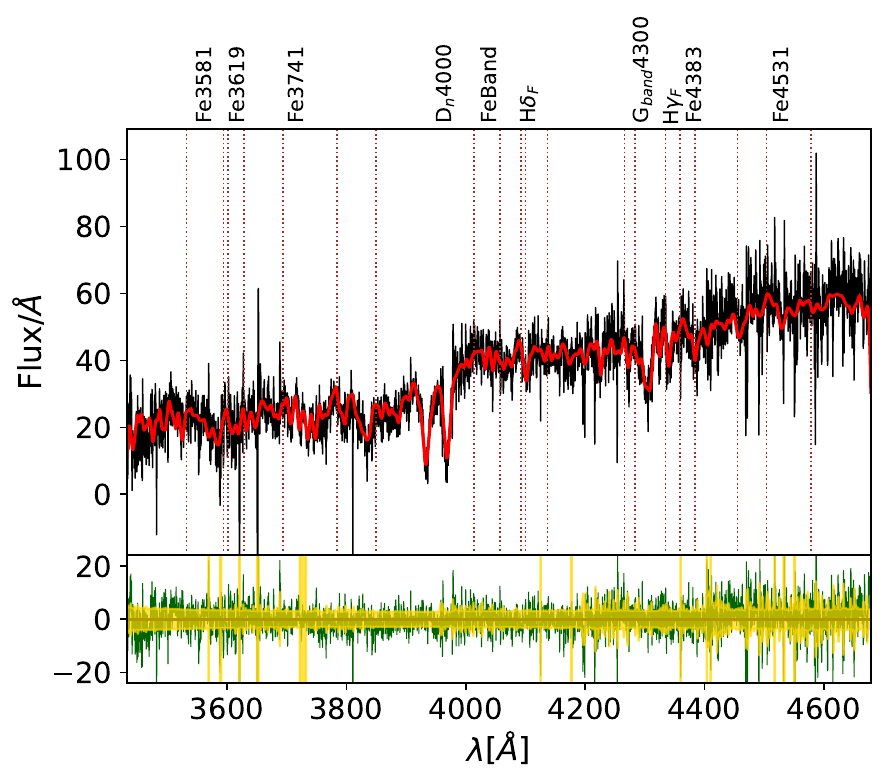}} \quad
        \subfloat[][\emph{}]
        {\includegraphics[width=0.45\textwidth]{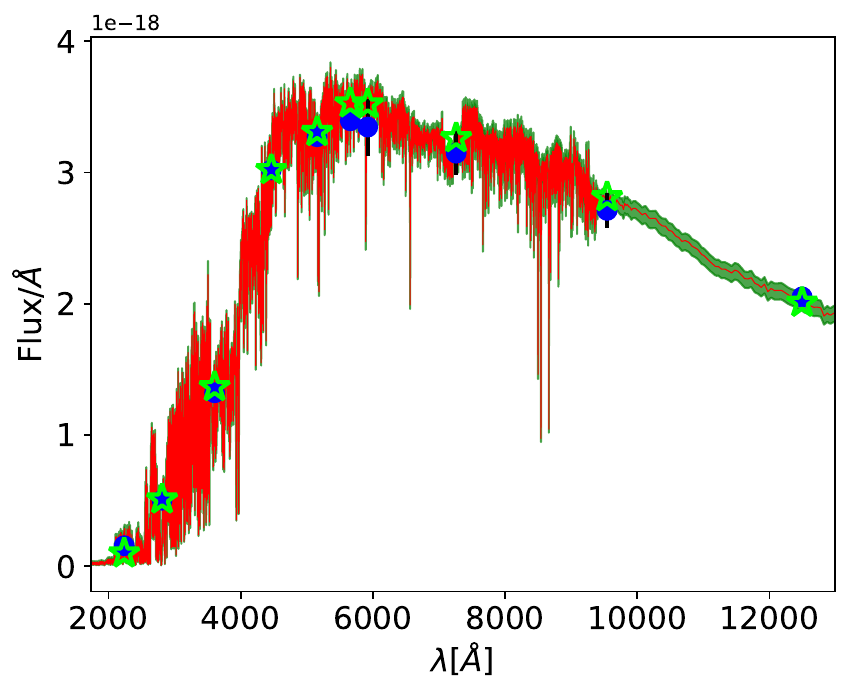}} \quad
 
   \caption{Panel (a), top: Example of the best fit (in red) applied to the entire LEGA-C spectrum of the LEGA-C ID 209466 galaxy (in black). Panel (a), bottom: Difference between the observed spectrum and the best fit, shown in green, and the relative uncertainties of the observed spectrum, shown in yellow. Panel (b): Observed photometric measurements (blue points), best-fitting template (in red), and photometric measurements on the best-fitting template (lime stars). The green shaded region indicates the $68\%$ confidence interval of our posterior.}
              \label{fig:example_specphot}
    \end{figure*}

Therefore, we computed the total posterior probability distribution using the likelihood given by $\mathcal{L} = e^{-\chi^2/2}$, with

\begin{equation}
\begin{split}
    \chi^2 =& \sum_{i}{\left(\frac{F_{obs_i}-F_{syn_i}}{\sigma_{obs_i}}\right)}^2 + {\left(\frac{D_\text{n}4000_{obs}-D_\text{n}4000_{syn}}{\sigma_{D_\text{n}4000_{obs}}}\right)}^2 +  \\ &+ \sum_{i}{\left(\frac{Phot_{obs_i}-Phot_{syn_i}}{\ln{f_x}*\sigma_{Phot_{obs_i}}}\right)}^2,
\end{split}
\end{equation}
where the first term is the contribution of the comparison using the FIF approach, F$_{syn_i}$ is the flux of the synthetic spectrum along the feature of each index, and F$_{obs_i}$ is the flux of the observed spectrum with the error $\sigma_{obs_i}$. The second term is the contribution of the D$_n4000$ comparison, where D$_\text{n}4000_{syn}$ is the index measurement on the synthetic template, and D$_\text{n}4000_{obs}$ is the index measurement on the observed spectrum with the error $\sigma_{D_\text{n}4000_{obs}}$. Finally, the last term is the contribution of the comparison with the photometric data, where Phot$_{syn_i}$ is the photometric flux in the i-th band of the synthetic spectrum, and Phot$_{obs_i}$ is the photometric flux in the i-th band of the observed galaxy with the error $\sigma_{Phot_{obs_i}}$. The photometric error is multiplied by $\ln f_{x}$, where $x$ can be $uv$, $opt$, or $ir$ as described above.
We derived posterior probability distributions and the Bayesian evidence using the nested sampling Monte Carlo algorithm
MLFriends \citep{2016S&C....26..383B,2019PASP..131j8005B} using the UltraNest\footnote{\url{https://johannesbuchner.github.io/UltraNest/}} package \citep{2021JOSS....6.3001B}. We assumed uniform prior for all the parameters considered, summarised in Table~\ref{table:params}. 
Figure~\ref{fig:corner_all} shows an example of the joint and marginal probability distributions for all the fitted parameters, while Fig.~\ref{fig:example_specphot} shows the best-fit applied to both the entire LEGA-C spectrum and the photometric data for a representative quiescent galaxy (LEGA-C ID = 209466). It is evident that our fit provides reasonable fits for both the spectrum and the photometric data of the galaxy.
\begin{table}
\caption{The seven free parameters fitted to our spectroscopic and photometric data, along with their associated prior distributions.}
\centering  
\begingroup
\begin{tabular}{llll}
\hline\hline      
Parameter & Unit & Range & Prior \\
\hline
 Age$_{\text{model}}$ & Gyr & (0, 8) & Uniform  \\
 $\text{[Z/H]}$ & dex & (-2.23, 0.55) & Uniform \\
 $\tau$ & Gyr & (0.1, 2) & Uniform \\
A$_V$ & mag & (0, 1) & Uniform  \\
$\ln{f_{uv}}$ & dex & (-2, 10) & Uniform \\
$\ln{f_{opt}}$ &dex & (-2, 10) & Uniform\\
$\ln{f_{ir}}$ &dex & (-2, 10) & Uniform \\
\hline

\end{tabular}
\endgroup
\label{table:params}
\end{table}

\subsection{Hierarchical Bayesian modelling}
\label{subsec:hierarc}
\begin{figure}
   \centering
   \includegraphics[width=0.5\textwidth]{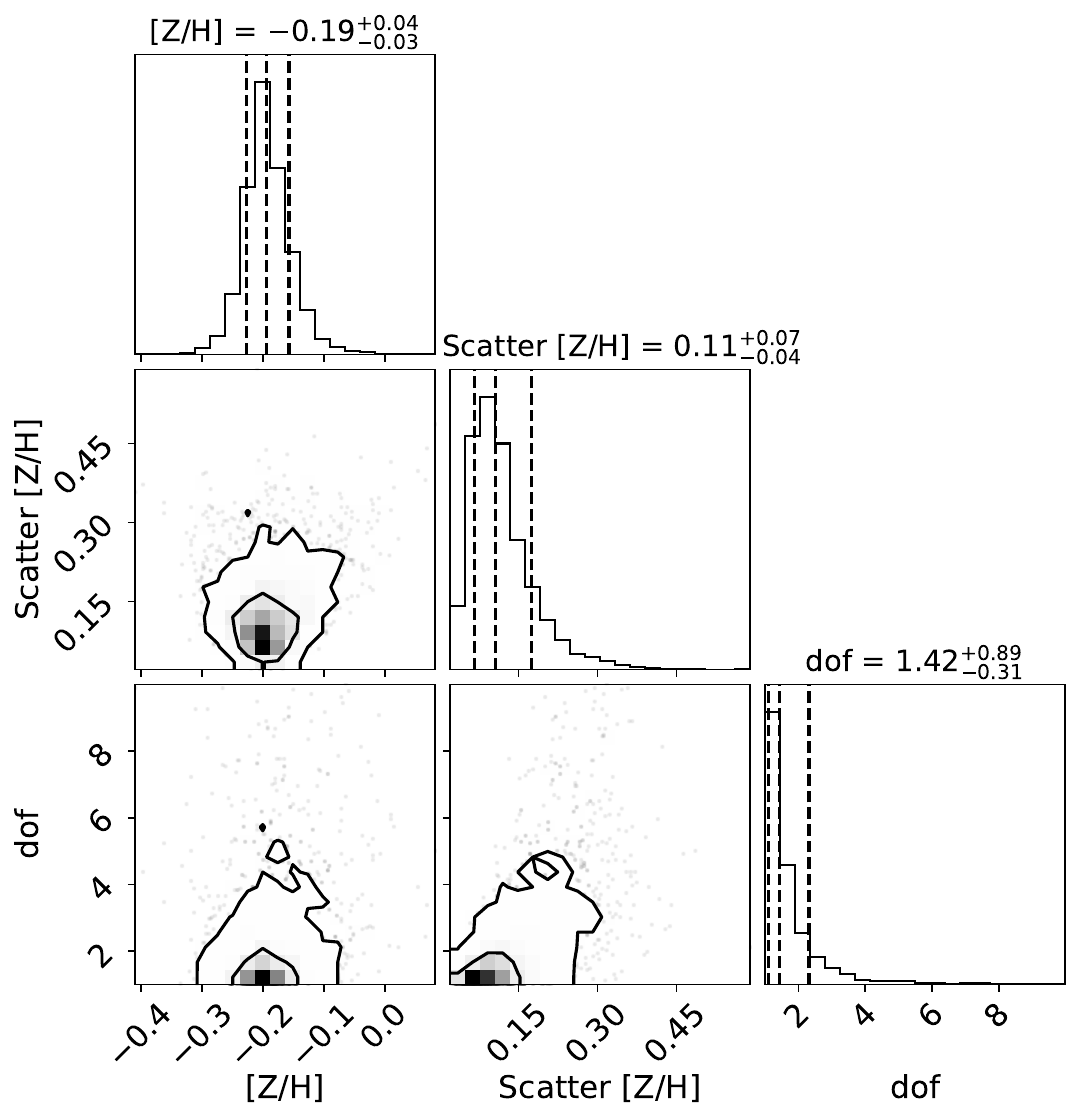}
      \caption{Corner plot summary of the posterior distributions of the [Z/H], its scatter, and the degrees of freedom of the Student-t distribution obtained for the galaxies in the X-ray group environment bin. Contours are at the $68\%$ and $95\%$ probability levels. The $16\%$, $50\%,$ and $84\%$ intervals are indicated by dashed lines.}
         \label{fig:corner_hier_zh_xray}
   \end{figure}

To assess the stellar population parameters in each environment bin, we combined the obtained results for each individual galaxy using the hierarchical Bayesian modelling. In the hierarchical framework, our models consist of two levels: the first level involves the measurements of the parameters for each individual galaxy, while the second level describes how the measurements are distributed within each environment bin. Differently from the classic stacked spectra fit,
this approach has several advantages as it avoids introducing  correlated noise caused by smoothing the individual spectra to a common velocity dispersion  and by the continuum interpolation with polynomials \citep[see Appendix B of][]{beverage2023}.
Following the `a posteriori' approach as in \cite{beverage2023}, as first level of the models we computed the posterior probability distribution of each physical parameter listed in Table~\ref{table:params} for each individual galaxy as detailed in Sect.~\ref{subsec:jointphotandspec}. Then, as a second level of modelling, we fitted the posterior of  each parameter  with a Student-t distribution in each environment bin. The use of the Student-t distribution accounts for potential high kurtosis in the parameter distributions. This approach provides a mean value for each parameter in each environment bin, the intrinsic scatter of our sample, along with a reliable estimate of the uncertainties. We assumed the same prior stated in Table~\ref{table:params} for each parameter, then we applied a logarithmically uniform prior for the intrinsic scatter between $0.01$ and $10$, and a logarithmically uniform prior for the degrees of freedom of the Student-t distribution between $1$ and $10$, where $1$ represents an heavy-tailed distribution and $10$ a gaussian one. Figure~\ref{fig:corner_hier_zh_xray} shows an example of the posterior probability distribution of the [Z/H] parameter, its scatter, and the degrees of freedom of the Student-t distribution for the X-ray groups environment bin.

\section{Results and discussion}
\label{sec:resultdisc}
As described in the previous sections, we spectroscopically selected $74$ massive quiescent galaxies in the COSMOS Wall volume, and we analysed their physical properties performing a joint fit of their spectra and photometric data. 

\begin{table*}
\caption{Median values of the marginalised posterior distribution of the age$_\text{model}$, mass-weighted age, [Z/H], $\tau$, and $A_V$ of the second level model.}            
\label{table:valerr}      
\centering          
\begin{tabular}{l c c c c c c}     
\hline\hline       
Environment bin & number of galaxies & Age$_{\text{model}}$ & Mass-weighted age & [Z/H] & $\tau$ & $A_V$ \\
& & [Gyr] & [Gyr] &[dex] & [Gyr] & [mag]   \\
\hline   \\                  
  X-ray groups & 16  & 5.83$^{+0.66}_{-0.63}$ & 5.52$^{+0.67}_{-0.64}$ &  -0.19$^{+0.04}_{-0.03}$ &  0.20$^{+0.06}_{-0.05}$ & 0.24$^{+0.03}_{-0.03}$ \\\\      
   Non-X-ray groups & 22  &5.36$^{+0.46}_{-0.47}$ & 4.62$^{+0.50}_{-0.45}$ &-0.17$^{+0.06}_{-0.06}$ & 0.50$^{+0.07}_{-0.06}$ & 0.28$^{+0.04}_{-0.04}$ \\\\
   Field & 36 & 4.77$^{+0.37}_{-0.36}$ & 4.04$^{+0.35}_{-0.37}$ & -0.02$^{+0.05}_{-0.05}$ & 0.50$^{+0.06}_{-0.06}$ & 0.30$^{+0.02}_{-0.02}$ \\\\
   All & 74 & 5.18$^{+0.25}_{-0.27}$ & 4.46$^{+0.25}_{-0.26}$ & -0.10$^{+0.04}_{-0.04}$ & 0.48$^{+0.04}_{-0.04}$ & 0.30$^{+0.02}_{-0.02}$
   \\\\
\hline                  
\end{tabular}
\tablefoot{The errors on the median values refer to the 16th$^{}$ and 84th$^{}$ percentiles.}
\end{table*}
We combined the results obtained on individual galaxies to characterise the galaxy properties for the entire sample and in the three different environments defined in Sect.~\ref{sec:wall}.
In the following we present the obtained results and compare them with other studies in literature (Sect.~\ref{sub:results}), and we finally discuss our results in the framework of the environmental effect on the evolution of the massive quiescent galaxies (Sect.~\ref{sub:effect}).

\subsection{Results and comparison with the literature}
\label{sub:results}

 As described in Sect~\ref{sec:wall}, our selected sample of massive quiescent galaxies is representative of the entire population of galaxies in the stellar mass range of  $10.47 < \log{M/M_\odot} < 11.4$. For each galaxy in the sample, we derived the following physical parameters: stellar mean metallicity, mean Age$_{\text{model}}$, star formation timescale ($\tau$), and dust extinction (A$_{V}$).  Using a Spearman rank correlation test, we found that none of these properties in each of the three environment bins are correlated with the stellar mass of the galaxies in each subsample with significance $>2\sigma$. Therefore, the chosen small mass dynamic range and the uniformity of the mass distributions across the different environments enable the study of the environmental effects on galaxy properties. We then combined the physical properties of the individual galaxies using a hierarchical Bayesian approach to determine the parameters for the entire sample and the characterising the three environments of different galaxy densities.

\begin{figure*}
   \centering
   \subfloat[][\emph{}]
        {\includegraphics[width=0.32\textwidth]{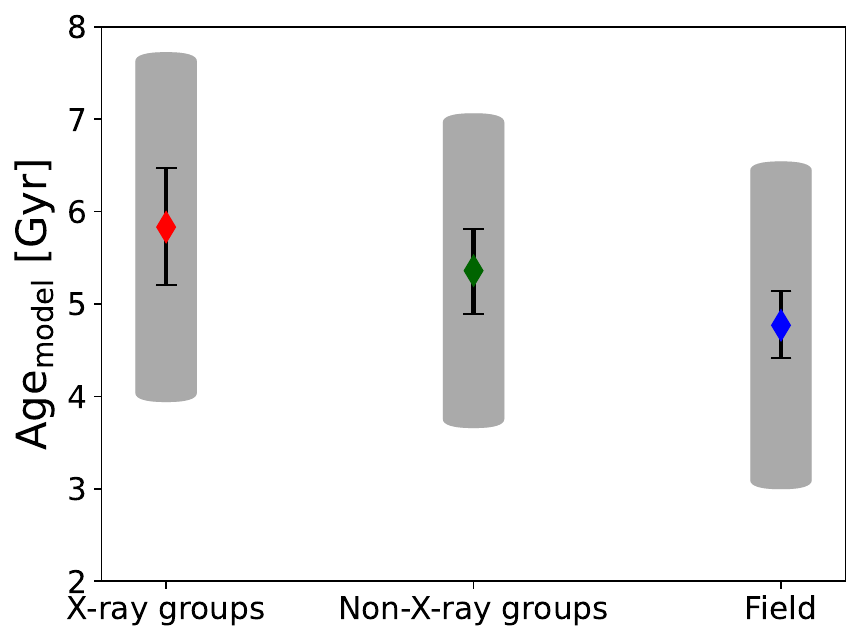}} \quad
        \subfloat[][\emph{}]
        {\includegraphics[width=0.32\textwidth]{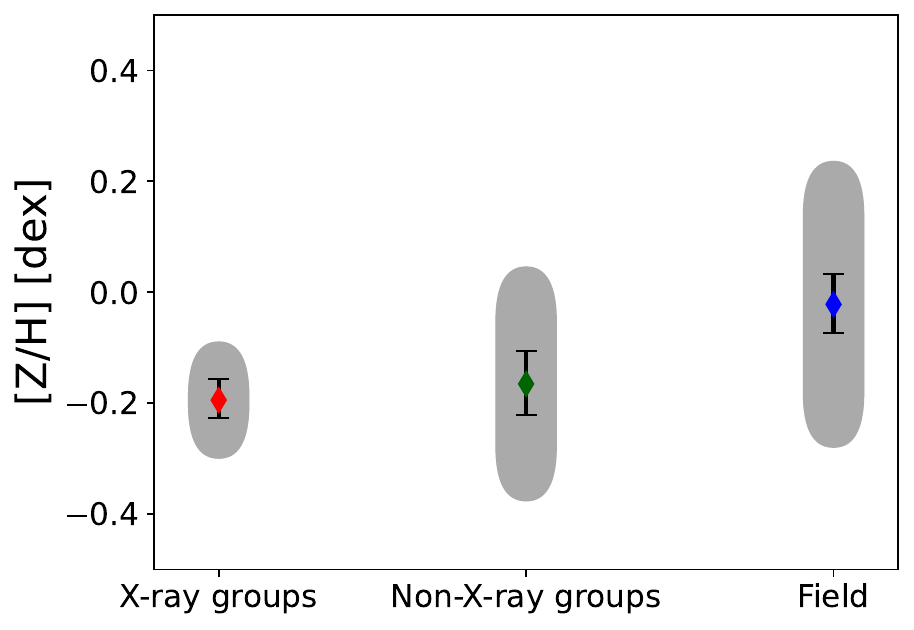}} \quad
        \subfloat[][\emph{}]
        {\includegraphics[width=0.32\textwidth]{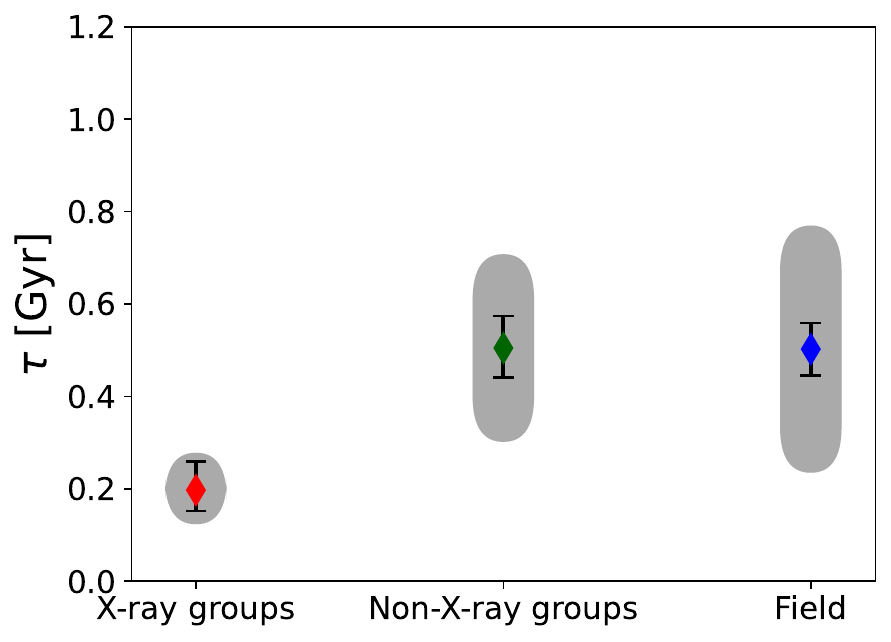}} \quad
 
   \caption{Age$_\text{model}$, [Z/H], and $\tau$ in different environments. Upper left panel: Median values of the age$_\text{model}$ with their relative error bars for different environments. Upper right panel: Median values of [Z/H] and their errors for different environments. Lower panel: Median values of $\tau$ and their errors for different environments. The grey shaded areas represent the intrinsic scatter of the individual posterior of the galaxies in each environment bin. }
              \label{fig:agezh_env}
    \end{figure*}

Figure~\ref{fig:agezh_env} shows the results obtained for Age$_{\text{model}}$, [Z/H] and $\tau$ as a function of the environment. The shaded grey regions represent the intrinsic scatter of the individual measured parameters in each environment bin. It is interesting to note that all the fits have shown values of Age$_{model}/\tau$ higher than $3$, confirming the purity and the fast star formation of the selected quiescent sample.
Regarding the A$_V$ parameter, we did not consider it as a quantity containing physical information, as it is has been derived from template fitting procedures and may partially account for potential mismatches between observations and models. At the same time, it is reassuring that we found a value of around $0.26$ mag across all the environments (see Table~\ref{table:valerr}) is consistent with results reported by other authors for massive galaxies at intermediate redshift \citep[e.g.][]{citro2016}.

From Table~\ref{table:valerr} and Fig.~\ref{fig:agezh_env}, a decreasing trend in the mass-weighted age of galaxies as a function of the environmental density is clearly evident. Galaxies in the X-ray groups have a typical mass-weighted mean age of $5.5$ Gyr, and they become progressively younger moving towards less dense environments, with a typical age of $4$ Gyr in the field, with a significance of almost $2 \sigma$. This means that galaxies in X-ray groups are about $1.5$ Gyr older than those in the field, although the intrinsic scatter measured across all three environments exceeds $1.5$ Gyr.

The top-right panel of Fig.~\ref{fig:agezh_env} shows that galaxies in the field exhibit a stellar metallicity value that is around $0.15$ dex higher than those in denser environments, with a significance greater than $2 \sigma$. Galaxies in both X-ray and non-X-ray groups display similar metallicity values, around [Z/H] = $-0.18$ (i.e. $60\%$ of the solar value).
 
Galaxies in X-ray groups appear quite homogeneous in their values of $\tau$, representing the star formation timescale, and  their measured $\tau$ value is significantly lower than those found in both the non-X-ray group environment and the field, even when accounting for the intrinsic scatter. It is worth noting that galaxies in non-X-ray groups and in the field present similar star formation timescales, but they differ by $0.6$ Gyr in their mean mass-weighted age.
The smaller intrinsic scatter observed in the stellar metallicity and $\tau$ for the sample in X-ray groups could indicate homogeneity in the physical processes that shaped these galaxies.
In the next subsection we discuss possible explanations for this finding.


To check the robustness of our results, we compared them with those found by other authors in the literature. In particular, 
\cite{darvish2017} proposed a different definition of the environment around the COSMOS Wall volume. In their work, they used the density field Hessian matrix to disentangle the cosmic web into clusters, filaments and field \citep[see][for full details]{darvish2016,darvish2017}. The main difference between the \cite{Iovino2016} and the \cite{darvish2017}  definitions of environment is that the former applied a group detection algorithm and then checked for X-ray emitting or non-X-ray emitting ones, considering all other galaxies located in the field, while the latter locates the galaxies based solely on the density of the environment, considering a global smoothing width of $0.5$ Mpc, thus somewhat larger than the typical virial radius of groups. 
\cite{darvish2017} performed a less detailed analysis of the galaxy stellar population properties in different environments compared to the one presented here, and found that the intrinsic colours of massive quiescent galaxies do not depend on the environment.
We considered the results obtained in our analysis while classifying galaxies following \cite{darvish2017} criteria.
Our results confirm the trend that galaxies in denser environments tend to be older, have lower stellar metallicities, and shorter formation timescales compared to those in the field.

We also compared our results with those of \cite{sobral2022}. Exploring the full LEGA-C sample at $0.6 < z < 1$ and using the \cite{darvish2017} definition of environment, they found that D$_n4000$ and H$\delta$ indices for quiescent galaxies vary with both stellar mass and local density. In particular, at fixed stellar mass, their results suggest that quiescent galaxies residing in higher-density regions are older, meaning they formed the bulk of their stars earlier and quenched earlier. Their results are fully consistent with what we found in our analysis. Indeed, even in the narrow bin of mass and redshift we considered, we found that quiescent galaxies are older and with a shorter formation timescale in denser environment than in the field.

We compared our global results with those from other studies using LEGA-C data. Within the same redshift and mass range, the mass-weighted age of our sample ($\sim 4.5$ Gyr) is consistent within 0.5 Gyr with previous studies \citep[e.g.][]{gallazzi2014charting,chauke2018star,beverage2023,kaushal2024}, despite differences in stellar population models and analysis methodologies. It is also worth noting that the mean stellar metallicity of our sample ($-0.1$ dex) matches with the [Fe/H] abundance measured by \cite{beverage2023}. This further supports the robustness of our measurements, based solely on Fe indicators, suggesting that our total metallicity estimate effectively serves as a proxy for the [Fe/H] abundance.

In the local Universe, many previous studies have analysed the possible dependence of the physical properties of galaxies on the environment in which they are located. \cite{sanchez2006medium} analysed quiescent galaxies in different environments, including Virgo and Coma clusters, finding that quiescent galaxies in lower-density environments appear younger by $1-2$ Gyr that those in denser environments, at given galaxy stellar mass. Their results is consistent with what we found in this work at higher redshift, suggesting an almost passive evolution from $z \sim 0.7$ to $z = 0$ for the most massive quiescent galaxies. A different conclusion is presented by \cite{thomas2010}. Using low-redshift data from the Sloan Digital Sky Survey ($0.05 \le z \le 0.06$), they found that the age and the stellar metallicity of the most massive quiescent galaxies do not depend on the environment and its density. Their results seem to suggest that the stellar population properties of the most massive quiescent galaxies are mainly driven by self-regulation processes related to intrinsic galaxy properties such as stellar mass. Our analysis of the properties of massive quiescent galaxies in the very narrow slice of redshift of the COSMOS Wall volume, at halfway of the Universe life, points out important differences related to the environment in which these galaxies are located. The differences we found are mainly in the mass-weighted age (i.e. galaxies in high-density environments are about $1.5$ Gyr older than those in the field) and in the star formation timescale, which is roughly half in high-density environments compared to the field.

A possible explanation for the different results we found at $z \sim 0.7$ compared to those reported in \cite{thomas2010}, could be related to the small age differences we detected between X-ray groups galaxies and field ones. Indeed, precise age estimation is much more challenging in the local Universe than at higher redshift, since the typical spectral indices sensitive to the age of the stellar content lose sensitivity for old ages. Age measurement errors of about $10\%$ in the local Universe correspond to uncertainties of approximately $1-2$ Gyr for ages around $10$ Gyr. These uncertainties are of the same order as the age differences  we estimated between galaxies in high- and low-density environments. Conversely, a limitation of our work is the relatively small sample size compared to those available at lower-z. Large surveys of quiescent galaxies at intermediate to high redshift will help reduce the statistical uncertainties affecting our work.

\subsection{The environmental effect on stellar population parameters}
\label{sub:effect}
The study of the age and the stellar metallicity provides valuable insights into the  framework that describes the formation and the subsequent evolution of galaxies. In particular, thanks to the narrow redshift slice of the COSMOS Wall volume, the selection of a narrow stellar mass range, and detailed environment information from \cite{Iovino2016}, we are able to investigate how the environment can affect the evolution of massive quiescent galaxies. 

   \begin{figure}
   \centering
   \includegraphics[width=0.5\textwidth]{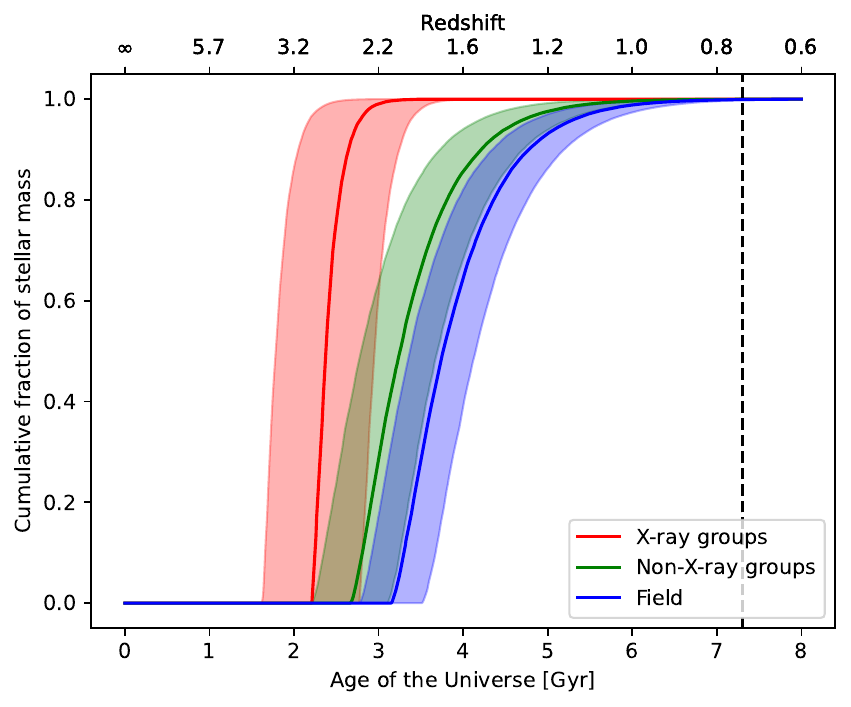}
      \caption{Cumulative mass fraction of the stellar mass as a function of the age of the Universe obtained for the galaxies in the three environment bins. The solid lines mark the $50\%$ percentile of the distributions, while contours are at the $16\%$ and $84\%$ confidence intervals. The vertical dashed line corresponds to the redshift of the COSMOS Wall structure ($z \sim 0.73$).}
         \label{fig:sfh}
   \end{figure}

Figure~\ref{fig:sfh} shows the cumulative mass fraction of the stellar mass as a function of the age of the Universe for galaxies in the three environments. As discussed in Sect.~\ref{sub:results}, X-ray group galaxies are older and exhibit a shorter star formation timescale  with respect to the less dense environments (see also Fig.~\ref{fig:agezh_env}). This suggests that they began forming their stars earlier and quenched at an earlier stage. These findings support the idea that the SFH of massive quiescent galaxies in denser environments is regulated by additional mechanisms compared to those acting in the field. Such mechanisms are responsible for the removal of the cold gas in galaxies, causing a sudden and rapid quenching of star formation in contrast to the slower quenching observed in field galaxies. The observed dependence of the quenching timescale on environmental density, combined with the uniformity in stellar mass among the studied galaxies,  suggests that mechanisms such as ram-pressure stripping, active predominantly in dense environments, play a key role in explaining the observed differences between galaxies in X-ray groups and those in the field.
This result is in good agreement with the SFHs derived for massive elliptical galaxies by \cite{guglielmo2015}, where the analysed galaxies show a faster decline of their SFR in clusters compared to the field. Interestingly, \cite{Zhou25} found a similar result, analysing the star-forming galaxies in the Wall volume. Their semi-analytical approach, applied to stacked spectra of star-forming galaxies in the Wall volume, grouped by bins of mass and environment, supports the view that galaxies in dense environments start their star formation earlier than field galaxies and have systematically shorter star formation timescales.

Regarding stellar metallicity, our analysis of massive quiescent galaxies shows that cluster galaxies have lower metallicity compared to those in the field. A lower stellar metallicity in high-density environments can be due to mechanisms such as ram-pressure stripping or tidal interactions, which remove gas reservoirs, curtail star formation, and limit the enrichment of the interstellar medium. Another possibility is that, in higher-density environments, the efficient dilution of metals in the cold gas reservoirs is driven by gas inflows from the lower metallicity intergalactic medium or by galaxy-galaxy interactions. In contrast, field galaxies benefit from prolonged gas availability, enabling extended star formation and chemical enrichment, which results in higher stellar metallicity. These results support the findings of \cite{calabro2022}, who reported lower metallicities in galaxies in high-density environments compared to the field, even at higher redshifts.
As mentioned above, it is important to note that, given the limited available wavelength range, we derived the total stellar metallicity in the sampled galaxies only on the basis of Fe indicators. In other words, the total metallicity we derived reflects the model assumption of the solar [Fe/Z] value.
Massive quiescent galaxies, which formed very rapidly and early in the cosmic time, may be characterised by a $[Fe/Z] < [Fe/Z]_\odot$ due to an enhanced abundance of $\alpha$ elements relative to Fe. Indeed, \cite{borghi2022} and \cite{bevacqua2023}, analysing the LEGA-C spectra of  quiescent galaxies found on average $[\alpha/Fe] > [\alpha/Fe]_\odot$. Consequently, the trend in metallicity shown in Fig.~\ref{fig:agezh_env} might reflect variations in $[\alpha/Fe]$ values across different environments, suggesting that the total metallicity could be higher than that measured. Notably, $\alpha$-enhancement is expected in systems where stars formed on short timescales \citep[][]{matteucci1987,thomas1999}. Indeed, as reported in Table~\ref{table:valerr}, we found that galaxies in X-ray groups are characterised by very short star formation timescales, shorter than those found in the other environments. This suggests that we may have underestimated the total metal content of galaxies in X-ray groups, which could be comparable to that of field galaxies, but with higher $\alpha-$enhancement in X-ray groups galaxies  than in the field ones. Again, this result well matches what \cite{Zhou25} found for the most massive star-forming galaxies in the Wall volume and their gas-phase metallicity.


The obtained results offer the opportunity to distinguish different evolutionary paths and quenching processes of massive quiescent galaxies. X-ray groups are characterised by hotter and denser intra-cluster medium compared to that in the non-X-ray groups \citep[e.g.][]{boselli2006,tonnensen2007}. As a result, galaxies in  X-ray groups are likely to be more significantly impacted by ram-pressure stripping than those located in small and less dense groups. Conversely, galaxies in less massive groups have a higher probability of merging and/or interacting with one another due their lower relative velocities. Figure~\ref{fig:agezh_env} shows that galaxies in non-X-ray groups exhibit younger ages than those in denser environment and older ages than those in the field. They have a stellar metallicity similar to that found in denser environment but a star formation timescale more akin to that of field galaxies. From Fig.~\ref{fig:sfh} we can see that these galaxies started their star formation activity at a similar epoch as those in denser clusters, but quenched later due to less efficient quenching mechanisms.
 While our results suggest a non-negligible role of dynamical processes affecting the gas supply available to galaxies in X-ray groups, the earlier formation time could also imply that star formation occurred in different internal physical conditions, leading to a more efficient star formation activity (and thus an earlier exhaustion of gas supply)  even in absence of an active role of the environment. In this respect, however, we can leverage the sample of the non-X-ray groups, in which galaxies formed earlier than in the field but with the same timescale, possibly due to a similar rate of galaxy interactions and mergers in these two environments. 
 
Higher S/N data covering a larger wavelength range will provide a clearer picture of the evolution of massive quiescent galaxies. Indeed, measuring the [$\alpha/Fe$] is essential for obtaining a reliable estimation of the total stellar metallicity of quiescent galaxies, enabling a deeper insight into their complex assembly history across different environments.


\section{Summary and conclusions}
\label{sec:conclusion}
We have presented an analysis of the stellar populations of a sample of 74 massive quiescent galaxies in the COSMOS Wall to investigate the role of the environment in galaxy evolution. The COSMOS Wall is a structure at $z \sim 0.73$ that contains a large variety of environments, from rich and dense clusters down to empty field regions. The analysed sample was derived using the LEGA-C spectroscopic data; we selected galaxies with no detectable [OIII]$_{\lambda5007}$ or H$\beta$ emission lines and at most a weak [OII]$_{\lambda3727}$ line. Additionally, we excluded post-starburst galaxies by requiring the H$\delta$
absorption line  with equivalent widths $< 3$ \ \AA. 
From this sample, we selected galaxies in a narrow range of mass, $10.47 <\log{M/M_\odot} < 11.4$, in order to have a representative, homogeneous sample in the slice of redshift of the COSMOS Wall across the different environments. We used the environment characterisation from \cite{Iovino2016}, from which we found $16$ galaxies in groups with X-ray detection, $22$ in groups without X-ray detection, and $36$ in the field.

We combined the LEGA-C survey spectroscopic data, which have high S/N and resolution suitable for studying the stellar population properties of galaxies, with the multi-wavelength photometric COSMOS2020 catalogue.
We performed a joint fitting of spectra and photometric data using nested sampling techniques in order to measure the mass-weighted age, stellar metallicity, and star formation timescale of individual galaxies. After that, we assessed the stellar population parameters of each environment bin using a hierarchical Bayesian approach.
Our findings can be summarised as follows:
\begin{itemize}
    \item Quiescent galaxies in denser environments are about $1.5$ Gyr older than those in the field.
    \item There is no  significant difference in stellar metallicity between galaxies in the X-ray and non-X-ray groups, both showing values of about $60\%$ of the solar value. Field galaxies exhibit metallicities approximately 
$0.15$ dex higher. 
    \item Galaxies in the X-ray groups have shorter star formation timescales. Those in the other two environment bins have similarly higher timescales despite a $1$ Gyr difference in their mass-weighted ages.
\end{itemize}
Our results support the idea that the SFH of massive quiescent galaxies in denser environments is regulated by mechanisms different from those acting in the field. In particular, the presence of a hot and dense intra-cluster medium in the X-ray groups strongly suggests that ram-pressure stripping plays a key role in the quenching process. Processes like cosmological starvation could explain the physical parameters  we observed in galaxies in the field, which have younger ages and longer star formation timescales compared to  those in the X-ray groups. 
The longer star formation timescale of galaxies in non-X-ray groups indicates a more complex evolutionary history, likely driven by more frequent interactions and mergers, which is  typical in less dense and smaller groups.

Upcoming surveys with new, high multiplexed, large field-of-view spectrographs, such as StePS at the WEAVE and 4MOST instruments \citep[][]{iovino2023,iovino2023Msn}, will provide the spectra of thousands of galaxies and cover wide spectral ranges with S/N suitable for measuring stellar population parameters. These data will allow us to measure the [$\alpha/Fe$] of the stellar populations, which is crucial for obtaining a clearer picture of the star formation timescale of the massive quiescent galaxies and further understanding their complex assembly history across various environments.

\begin{acknowledgements}
We thank the anonymous referee for carefully reading the manuscript and for the stimulating suggestions that helped to improve the paper.
S.Z., A.I., M.L., S.B., M.B., O.C., L.P., D.V., E.Z. and F.D. acknowledge financial support from INAF Large Grant 2022, FFO 1.05.01.86.16, and the Italian Ministry grant Premiale MITIC 2017. The authors wish to acknowledge the generous support of ESO staff during service observations of program ESO 085.A-0664 (The COSMOS Structure at z = 0.73: Exploring the Onset of Environment-Driven Trends), which provided the valuable environmental characterisation on which this work is based.
\end{acknowledgements}

%
%

\bibliographystyle{aa}
\bibliography{aanda}

\begin{appendix} 

\label{sec:appA}

\end{appendix}

\end{document}